\documentclass[english]{aa}
\usepackage[T1]{fontenc}
\usepackage{geometry}
\usepackage{bm}
\usepackage{amstext}
\usepackage{amsmath}
\usepackage{graphicx}
\usepackage[authoryear,round]{natbib}
\usepackage{array}
\usepackage{multirow}

\makeatletter

\providecommand{\tabularnewline}{\\}

\makeatother

\usepackage{babel}
\begin{document}
\global\long\def\grad{\bm{\nabla}}

\title{Iterative inversion of synthetic travel times successful at recovering sub-surface profiles of supergranular flows}

\author{\author{Jishnu Bhattacharya\inst{1} \and Shravan M. Hanasoge \inst{1}\and Aaron C. Birch\inst{2} \and Laurent Gizon \inst{2,3,4}}
\institute{Tata Institute of Fundamental Research, Mumbai, India \and Max Planck Institute for Solar System Research, Justus-von-Liebig-Weg 3, 37077 G\"ottingen, Germany\and 
Institut f\"ur Astrophysik, Georg-August-Universit\"at G\"ottingen, 37077 G\"ottingen, Germany \and
Center for Space Science, NYUAD Institute, New York University Abu Dhabi, Abu Dhabi, UAE}}

\titlerunning{Parametrized inversion strategy to obtain supergranular flows}
\authorrunning{Bhattacharya et al.} 

\abstract{}{We develop a helioseismic inversion algorithm that can be used to recover sub-surface vertical profiles of 2-dimensional supergranular flows from surface measurements of synthetic wave travel times.}
{We carry out seismic wave-propagation simulations through a 2-dimensional section of a flow profile that resembles an averaged supergranule, and a starting model that has flows  only at the surface. We assume that the wave measurements are entirely without realization noise for the purpose of our test. We expand the vertical profile of the supergranule stream function on a basis of B-splines. We iteratively update the B-spline coefficients of the supergranule model to reduce the travel-times differences observed between the two simulations. We carry out the exercise for four different vertical profiles peaking at different depths below the solar surface.}
{We are able to accurately recover depth profiles of four supergranule models at depths up to $8-10\,\text{Mm}$ below the solar surface
using $f-p_4$ modes, under the assumption that there is no realization noise. We are able to obtain the peak depth and the depth of the return flow for each model.}
{A basis-resolved inversion performs significantly better than one where the flow field is inverted for at each point in the radial grid. 
This is an encouraging result and might act as a guide in developing more realistic inversion strategies that can be applied
to supergranular flows in the Sun.}

\date{}
\maketitle

\section{Introduction}

Convective flows on the solar surface exhibit several length
scales \citep{Nordlund2009}. Small-scale granules ($\sim 1\,\text{Mm}$) are well studied and characterized. The sub-surface profile and
physics behind relatively larger scale flows have eluded a convincing
explanation. Supergranules --- believed to be overturning
flows spanning $35$ Mm horizontally on average \citep{Hathaway2000,Rieutord2008}
--- remain one of the fronts where sub-surface imaging
has had limited success. Sketching a complete picture of a supergranule
involves understanding the nature and magnitude of upflows near the
cell center, downflows at the edges of the cell, horizontally diverging
flows from the center of the cell towards the edge, as well as deeper
return flows that might be present. There are several techniques used
to study sub-surface flows \citep[see][and references therein]{Gizon2010},
out of which we shall focus on time-distance seismology \citep{Duvall1993}.
This approach lets us relate travel-time shifts of seismic waves in
the sun to surface and sub-surface velocities, thereby setting
up an inverse problem where measurements of wave travel-times can
be used to estimate flow fields in the solar interior. Time-distance
seismology has been widely used to recover flows in the Sun 
\citep{Duvall2000,Zhao2003,Zhao2004,Jackiewicz2008,Duvall2012,Svanda2012}, however when applied specifically to supergranules, the sub-surface flow profiles have been hard to pin down. 

Despite successful measurements of supergranular flows on the solar surface \citep{Rieutord2008,Duvall2010,Svanda2013}, seismic studies have not been successful at consistently reproducing the sub-surface profiles of supergranules. \citet{Duvall1998} used correlations between inverted
surface and deeper flows to determine the lower bound of the supergranule pattern where the convective cell overturns, and obtained a depth of $8\,\text{Mm}$.
\citet{Zhao2003} used time-distance seismology to invert MDI data
and inferred that the depth of a supergranule was $15\,\text{Mm}$.
\citet{Braun2004} applied phase-sensitive holography to MDI data
and concluded that detection of return flow below $10\,\text{Mm}$
would prove to be a significant challenge due to contamination from
neighboring supergranules. \citet{Woodard2007} used Fourier-space
correlations in the observed wave field, and found power to extend
down to $6\,\text{Mm}$ below the surface before noise took over; similar results were also obtained by \citet{Braun2007} using helioseismic holography and \citet{Jackiewicz2008} using time-distance seismology. Using
measurements of horizontal flow divergences and vertical velocity
obtained from Solar Optical Telescope (SOT) on board the Hinode satellite,
\citet{Rieutord2010} estimated the vertical scale height of supergranules
to be $1\,\text{Mm}$. \citet{Duvall2012} used a ray-theoretic forward
modeling approach to fit center-annulus travel-time differences obtained
using Gaussian models of vertical velocity to those obtained from
a kinematic model of an ``averaged supergranule'' derived using Dopplergrams
from Helioseismic and Magnetic Imager \citep[HMI:][]{Schou2012} on-board
the Solar Dynamics Observatory (SDO) spacecraft. They estimated that
the depth corresponding to peak vertical velocity to be $2.3\pm0.9\,\text{Mm}$,
where the value signified by $\pm$ represents the width of the model.
Additional evidence for a shallow supergranule was obtained by \citet{Duvall2014}.

The prevalence of disparate values and the dependence of results on
the specific technique being used calls for validation tests of seismic inversion algorithms. \citet{Dombroski2013} tested regularized least square inversions using helioseismic holography measurements for a supergranulation-like flow, but found that the inferred vertical flow has significant errors throughout the computational domain.  \citet{Svanda2011}
used subtractive optimally localized averages (SOLA) \citep{Jackiewicz2008}
and were able to recover 3-dimensional velocity fields from forward-modeled
travel-time maps generated from simulations of solar-like convective
flows. However \citet{Svanda2015} showed that reconstructed velocity
fields do not produce wave travel times that match the observed ones.
\citet{Degrave2014} tried validating time-distance SOLA inversions
using realistic solar simulations, and found that they could recover
horizontal flows till $5\,\text{Mm}$ below the solar surface, but
were unable to infer vertical flows accurately like \citet{Svanda2011}.
The authors attributed this to differences in measurement and analysis
techniques. A different approach was tried by \citet{Hanasoge2014}
and \citet{Bhattacharya2016}, who used full-waveform inversion \citep{Tromp2010}
to iteratively update a Cartesian 2-dimensional flow profile to minimize
travel-times misfit computed with respect to a model similar
to the average supergranule from \citet{Duvall2010}. They found that
seismic waves in their simulations were primarily sensitive to flow
updates close to the solar surface, and inversions focused on updating
these layers at the expense of deeper layers. This negative result
--- especially for a noise-free inversion --- was surprising in light
of the previous studies by \citet{Svanda2011} and \citet{Degrave2014},
and urged one to probe deeper into the reasons behind this mismatch.

In this work we follow an approach similar to \citet{Bhattacharya2016},
but ask an important question --- can we pose the problem
differently to avoid the interplay between the large number of parameters
being inverted for, and reduce the question to the fundamental one
of seismic sensitivity to flows? The means by which we approach the
question is to consider the pedagogic exercise of inverting for a 2-dimensional
section of the averaged supergranule of \citet{Duvall2012}, using
seismic waves that are are excited by sources located at specific
spatial locations. While the setup is not directly comparable with
solar observations, this serves as a computationally efficient starting
point to validate full-waveform inversion applied to the Sun. We project the supergranular flow model in a B-spline basis and solve an optimization problem to obtain the spline coefficients. We show that we are able to accurately recover the vertical profile of the averaged supergranule down to $8-10\,\text{Mm}$ below the surface. This result might help to guide the construction of improved inversion strategies to study the subsurface profile of an averaged supergranule in the Sun.

\section{Supergranule model}

We consider kinematic models of temporally stationary supergranules
in Cartesian coordinates. The entire analysis is two-dimensional primarily
for computational ease, although it provides us with an extra simplification
that would be absent in three dimensions --- that of a unidirectional
stream function. We choose coordinates $\mathbf{x}=\left(x,z\right)$,
where $z$ is a vertical coordinate that increases in the direction opposite to gravity, and $x$ denotes a horizontal direction with periodic boundary conditions. We set $z=0$ at the solar surface, therefore negative values of $z$ indicate depths below and positive values indicate heights above it.

A model of a supergranule can be described by its velocity field $\mathbf{v}\left(\mathbf{x}\right)$
that is embedded in a steady solar background characterized by a one-dimensional
density profile $\rho\left(z\right)$, sound-speed profile $c\left(z\right)$,
pressure $p\left(z\right)$, acceleration due to gravity $\mathbf{g}\left(z\right)=-g\left(z\right)\,\mathbf{e}_{z}$.
The velocity profile of the supergranule is chosen to resemble a section
through the averaged supergranule of \citet{Duvall2012}, differences
arising because of the computation being in Cartesian coordinates
rather than cylindrical. 

We enforce mass-conservation 
\begin{equation}
\grad\cdot\left(\rho\mathbf{v}\right)=0,
\end{equation}
and derive the velocity field from a stream function $\psi\left(\mathbf{x}\right)\mathbf{e}_{y}$ as \begin{equation}
\mathbf{v}=\frac{1}{\rho}\grad\times\left[\rho c\psi\mathbf{e}_{y}\right].\label{eq:v_psi}
\end{equation}
Our model for the supergranule stream function is 
\begin{eqnarray}
\psi\left(\mathbf{x}\right) & = & \frac{v_{0}}{c\left(z\right)}\frac{\mathrm{sign}\left(x\right)}{k}\,J_{1}\left(k\left|x\right|\right)\exp\left(-\frac{\left|x\right|}{R}\right)\times\nonumber \\
 &  & \exp\left(-\frac{\left(z-z_{0}\right)^{2}}{2\sigma_{z}^{2}}\right),\label{eq:superganule_psi}
\end{eqnarray}
where $J_{1}$ is the Bessel function of order $1$. The supergranule
stream function is zero at the cell center, peaks at a certain distance
away from the center before falling to zero and reversing sign; the
reversal in sign indicates a transition in the vertical velocity from
upflows to downflows. The model is highly simplified compared to supergranules
as observed on the solar surface, as it ignores the impact of magnetic
fields and other observed anomalous characteristics such as wave-like
nature associated with supergranules \citep{Gizon2003} and east-west
travel-time asymmetries \citep{Langfellner2015}. We fix the horizontal
length scales to $R=15\,\text{Mm}$ and $k=2\pi/\left(30\,\text{Mm}\right)$,
and use several combinations of parameters to characterize the vertical
profile. These parameters are listed in Table \ref{tab:vertical_params}. 

The flow field that we obtain from this stream function is 
\begin{eqnarray}
v_{x} & = & v_{0}\frac{\mathrm{sign}\left(x\right)}{k}\,J_{1}\left(k\left|x\right|\right)\exp\left(-\frac{\left|x\right|}{R}\right)\nonumber \\
 &  & \times\left(\frac{\left(z-z_{0}\right)}{\sigma_{z}^{2}}-\frac{\rho^{\prime}\left(z\right)}{\rho\left(z\right)}\right)\exp\left(-\frac{\left(z-z_{0}\right)^{2}}{2\sigma_{z}^{2}}\right),\\
v_{z} & = & v_{0}\left(\frac{1}{2}\left(J_{0}\left(k\left|x\right|\right)-J_{2}\left(k\left|x\right|\right)\right)-\frac{1}{kR}J_{1}\left(k\left|x\right|\right)\right)\nonumber \\
 &  & \times\exp\left(-\frac{\left|x\right|}{R}\right)\exp\left(-\frac{\left(z-z_{0}\right)^{2}}{2\sigma_{z}^{2}}\right).
\end{eqnarray}
We list the magnitude of the peak and surface velocities for these
flow fields in Table \ref{tab:Flow-velocity-magnitudes}. In subsequent
analysis, we shall refer to this velocity field with the superscript
``$\text{true}$'', ie. as $\mathbf{v}^{\text{true}}$, and similarly
for its components, to distinguish it from the flow velocity in the
iteratively updated flow model. We shall apply superscripts ``$\text{true}$''
and ``$\text{iter}$'' to other parameters wherever necessary, indicating
which model they correspond to. 

\begin{table}
\caption{\label{tab:vertical_params}Stream-function parameters}
\renewcommand{\arraystretch}{1.3}

\begin{tabular}{c c c c}
\hline 
Model & $z_{0}$ & $\sigma$  & $v_{0}$ \tabularnewline
& [Mm] & [Mm] & [$\text{m}/\text{s}$] \tabularnewline
\hline 
\hline 
SG1 & $-2.3$ & $0.9$ & $240$\tabularnewline
\hline 
SG2 & $-4$ & $1.6$ & $270$\tabularnewline
\hline 
SG3 & $-6$ & $2.2$ & $600$\tabularnewline
\hline 
SG4 & $-8$ & $2.8$ & $700$\tabularnewline
\hline 
\end{tabular}
\end{table}

\begin{table*}
\caption{Peak velocities, surface velocities and depths for the four models considered.
\label{tab:Flow-velocity-magnitudes}}
\renewcommand{\arraystretch}{1.3}
\begin{tabular}{c c c c c c c}
\hline 
Model & Max $v_{x}$  & Max $v_{x}$ at surface & $v_{x}$ peak depth  & Max $v_{z}$ & Max $v_{z}$ at surface & $v_{z}$ peak depth \tabularnewline
& [$\text{m}/\text{s}$] & [$\text{m}/\text{s}$] & [Mm] & [$\text{m}/\text{s}$] & [$\text{m}/\text{s}$] & [Mm]
\tabularnewline
\hline 
\hline 
SG1 & $601$ & $222$ & $1.4$ & $111$ & $5$ & $2.3$\tabularnewline
\hline 
SG2 & $390$ & $245$ & $2.7$ & $128$ & $6$ & $3.9$\tabularnewline
\hline 
SG3 & $614$ & $293$ & $4.2$ & $284$ & $7$ & $6.1$\tabularnewline
\hline 
SG4 & $539$ & $228$ & $5.2$ & $333$ & $6$ & $8.1$\tabularnewline
\hline 
\end{tabular}

\end{table*}

\section{Inversion Setup\label{sec:inv_setup}}

Waves in the Sun are driven near the solar surface by turbulent convection
associated with granules, with most of the excitation taking place
within $500\,\text{km}$ of the photosphere \citep{Stein2001}. Once
generated, seismic waves propagate under the restoring forces applied
by fluid pressure gradients and gravity. The wave displacement $\bm{\xi}(\mathbf{x},t)$ evolves
according to the equation
\begin{eqnarray}
\rho\partial_{t}^{2}\bm{\xi}+2\rho\mathbf{v}\cdot\grad\partial_{t}\bm{\xi} & = & \grad\left(c^{2}\rho\grad\cdot\bm{\xi}+\bm{\xi}\cdot\grad p\right)\nonumber \\
 &  & +\mathbf{g}\grad\cdot\left(\rho\bm{\xi}\right)+\mathbf{S},\label{eq:wave_equation}
\end{eqnarray}
where $\mathbf{S}$ represents sources that are exciting waves in the Sun. In our
simulation, we choose eight sources located at different horizontal
positions at a depth of $150\,\text{km}$ below the surface. The sources
represent ``master pixels'' \citep{Tromp2010,Hanasoge2011}, that
is their locations are chosen so that the emanating waves sample the
supergranule adequately. Each source fires independent of the others,
and produces waves that illuminate slightly different regions in the
Sun. For both the true supergranule and the iterated one, therefore,
we have eight different simulations that are computed in parallel,
each of which runs for $4$ hours in solar time. The simulation box
spans $800\,\text{Mm}$ horizontally over $512$ pixels, and extending
from $137\,\text{Mm}$ below the surface to $1.18\,\text{Mm}$ above
it vertically, resolved using $300$ pixels spaced uniformly in acoustic
distance. We place perfectly matched layers along the vertical boundaries
to absorb waves effectively.

We use the seismic wave propagation code SPARC \citep{Hanasoge2007}
to solve the wave equation in a convectively stabilized version of Model S \citep{ChristensenDalsgaard1996}. The model is one-dimensional and satisfies hydrostatic balance, and is stabilized by patching an isothermal layer to Model S above $0.98\,R_\odot$ \citep{Hanasoge2006}. The code SPARC computes seismic
wave fields by solving Equation (\ref{eq:wave_equation})
in the time domain using a low-dispersion and low-dissipation five-stage
Runge-Kutta time-stepping scheme \citep{Hu96}. Spatial derivatives
are computed using a sixth-order compact finite-difference scheme \citep{Lele92}
in the vertical direction, and using Fourier decomposition in the
horizontal direction. 

\begin{table}
\caption{\label{tab:receivers}Source-receiver distances}
\renewcommand{\arraystretch}{1.5}

 \begin{tabular}{ c c }
 \hline
  Radial Order & Receiver Distance  \tabularnewline
  & [$\text{Mm}$] \tabularnewline
  \hline\hline
  $f$ & $12-100$ \\ \hline
  $p_1$ & $12-120$ \\ \hline
  $p_2$ & $12-150$ \\ \hline
  $p_3$ & $12-200$ \\ \hline
  $p_4$ & $12-250$ \\ 
  
  \hline
  
 \end{tabular}

\end{table}

We apply ridge-filters to study the propagation of wavepackets corresponding to individual radial orders. The ideal filtering technique has been a subject of some debate: it was found by \citet{Svanda2013ApJ...775....7S} that inversions using a ridge-filtered approach produces results consistent with one using a phase-speed filtered approach, while \citet{Degrave2014} found that inversions using ridge-filtered travel times do not compare favorably with phase-speed-filtered ones, however we do not address this issue in the present work. Following \citet{Jackiewicz2008}, we filter the data by multiplying the wave spectrum by a function of the form $F_n\left(\nu,k\right)$, where $\nu$ represents temporal frequency, $k$ represents spatial frequency and $n$ represents the radial order. The filter function is constructed by separating modes corresponding to different radial orders using fourth-order polynomials of $k$. For each radial order, the spectral area enclosed between two such polynomials $\nu_{low}\left(k\right)$ and $\nu_{high}\left(k\right)$ is entirely included. We list the polynomials used for each radial order in Table \ref{tab:receivers}. The selected temporal-frequency band at each pixel in $k$ is terminated with a quarter of a cosine function over two pixels on both the high and low edges to ensure a smooth fall-off. Additionally low temporal frequency modes below $1.1\,\text{mHz}$ are filtered out to remove contributions from weak $g-$modes that arise as an artifact of the convectively stabilized background.

\begin{figure*}

\includegraphics[scale=0.67]{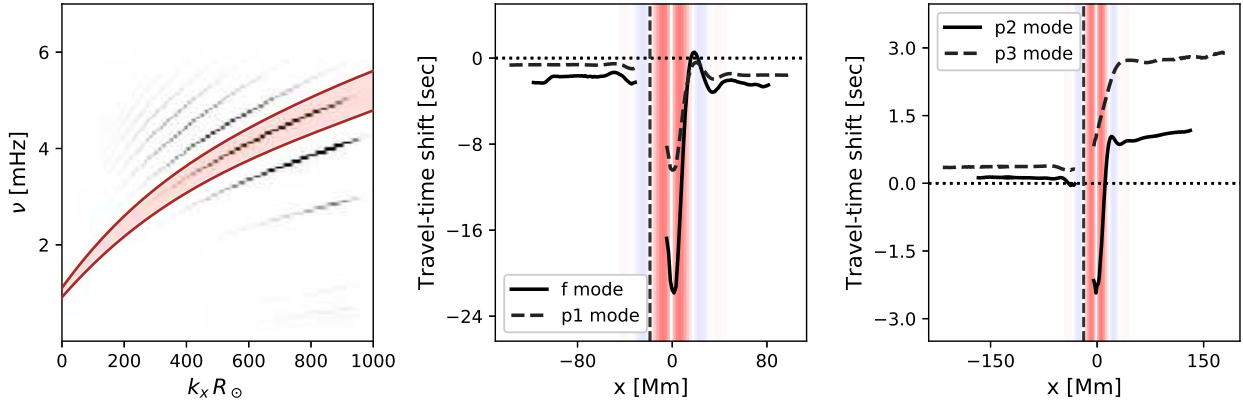}
\par

\caption{\label{fig:tt_modes}Left panel: Wave spectrum overlain with the function used to filter waves corresponding to the radial order $p_2$. Middle and right panels: Travel-time shifts between waves in the starting and true flow models for supergranule SG2, computed at all receivers for different radial orders. The horizontal location of the source is marked by the dashed vertical line. The horizontal flow field of the supergranule is indicated by the colored patch in the background, where red indicates outflows away from the cell center, and blue indicates inflows towards the cell center.}
\end{figure*}

We choose a group of pixels $200\,\text{km}$ above the surface and mark them as receivers. We list the horizontal locations of receivers for various radial orders in Table \ref{tab:receivers}. The wide range of receiver locations combines both short and large-distance measurements, thereby utilizing waves that probe various depths beneath the solar surface. We record filtered waveforms with time for $f$, $p_{1}$, $p_{2}$ and $p_{3}$ ridges at each receiver pixel for each simulation, and for the $p_{4}$ ridge as
well for the case of SG4.  An example of a spectrum along with a filter to extract waves corresponding to the radial order $p_2$, as well as measured travel-times at receivers for the model SG2 is depicted in Fig \ref{fig:tt_modes}. We define the travel-time misfit
\begin{equation}
\chi=\frac{1}{2}\sum_{\text{s}}\sum_{\text{ridge}}\sum_{\text{r}} 
\left(\tau_{s,r,ridge}^{\text{true}}-\tau_{s,r,ridge}^{\text{iter}}\right)^{2},\label{eq:travel-time-misfit}
\end{equation}
where $\tau_{s,r,ridge}^{\text{true}}$ refers to the waves emanating from 
the source $s$ whose ridge-filtered travel-time is measured
at the $r$-th receiver in presence of the true supergranule,
$\tau_{s,r,ridge}^{\text{iter}}$ is the travel-time measured in presence
of the iterated model for the same source-receiver locations and the same filter,
and the sum extends over source-receiver pairs as well as
different radial orders. We compute travel-times in a manner similar
to \citet{Gizon2002}, we describe the technique in detail in Appendix \ref{ttsec}.

Non-linear iterative time-distance inversions, as formulated in the
context of helioseismology by \citet{Hanasoge2014}, revolves around
reducing the misfit defined in Equation (\ref{eq:travel-time-misfit})
by sequentially improving a model of the supergranule. The scheme proceeds
by relating the travel-time misfit to an update in the supergranule
stream function through an integral relation as 
\begin{equation}
\delta\chi=\int_{\odot}d\mathbf{x}\,K_{\psi}\left(\mathbf{x}\right)\delta\psi\left(\mathbf{x}\right),\label{eq:kernel_psi}
\end{equation}
where $K_{\psi}\left(\mathbf{x}\right)$ is the kernel whose value
at any spatial point indicates sensitivity of wave travel-times to
local updates $\delta\psi\left(\mathbf{x}\right)$ in the supergranule
model. We use the adjoint source technique \citep{Hanasoge2011} to
compute the finite-frequency kernel $K_{\psi}\left(\mathbf{x}\right)$.
The steps involved in computing this kernel have been detailed in \citet{Hanasoge2014}
and \citet{Bhattacharya2016}.

\subsection{Basis-resolved Inversion}

Previous attempts at inversions by \citet{Hanasoge2014} and \citet{Bhattacharya2016}
have focused on solving for the stream function at each spatial location.
The number of parameters being inverted for was equal to the number
of spatial grid points, ie. for a $512\times300$ grid we would have
had $153600$ parameters. One of the questions we ask in this paper
is whether the large size of the parameter space had kept the previous
attempts from converging to the correct model. To answer this question,
we pose the inverse problem differently and make the following assumptions:
\begin{enumerate}
\item The supergranule stream function is separable in $x$ and $z$. This
assumption is made keeping in mind that we are primarily interested
in the depth of supergranules.
\item The value of the stream function at the surface and above is known.
The justification is that the layers at and above the surface are directly observed, and hence the flow velocities measured. 
\citep{Gizon2000,Rieutord2008,Duvall2010}
\end{enumerate}
A consequence is that the horizontal profile of the stream function is assumed to be known everywhere, and we only solve for its vertical profile. We express the stream function as
\begin{equation}
\psi^{\text{iter}}\left(\mathbf{x}\right)=f\left(x\right)g^{\text{iter}}\left(z\right),\label{eq:psi_separable}
\end{equation}
where $f\left(x\right)$ is entirely determined, and $g^{\text{iter}}\left(z\right)$ is known for $z>0$. We represent the true model for the stream function from Equation \eqref{eq:superganule_psi} in a similar manner as $\psi^\text{true}(\mathbf{x})=f(x)g^\text{true}(z)$, where $g^\text{true}(z)$ is the function that we seek to recover through the inversion.

We reduce the parameter space further by expanding the vertical profile of $\psi^\text{true}(\mathbf{x})$ in a basis of B-splines and inverting for the coefficients close to the surface. We expand the vertical profile $g^\text{true}\left(z\right)$ in a basis of quadratic B-splines with a given set of knots $\left\{t\right\}$ as
\begin{equation}
g^{\text{true}}\left(z\right)\approx\sum_{i=0}^{N-1}\beta^\text{true}_{i}\,B_{i}\left(z;t,k=2\right),\label{eq:basis_coeffs_all}
\end{equation}
where $\beta_{i}$ represent the B-spline coefficients, and $k=2$ indicates quadratic splines. The B-splines are ordered such that the index $i=0$ corresponds to the B-spline function that peaks the deepest, while the index $i=N-1$ corresponds to the one that peaks close to the upper boundary of our computational domain. The approximate equality is to be understood as the best fit in a least-square sense, since we choose a set of smoothing splines instead of interpolating ones. We describe the spline expansion in detail in Appendix \ref{app:spline}. We would like to point out that such an approach is applicable only to an ensemble averaged model of a supergranule, where the flow profile is expected to be smooth and representable using relatively few splines.

We split the coefficients into
two groups --- those above the surface and those below the surface.
Assuming that the coefficient with index $m$ corresponds to the B-spline
function that peaks at the solar surface, we rewrite Equation (\ref{eq:basis_coeffs_all})
as 
\begin{eqnarray}
g^{\text{true}}\left(z\right) & \approx & \sum_{i=0}^{m-1}\beta_{i}^{\text{true}}\,B_{i}\left(z;t,k=2\right)\nonumber \\
 &  & +\sum_{i=m}^{N-1}\beta_{i}^{\text{true}}\,B_{i}\left(z;t,k=2\right)\nonumber \\
 & = & \sum_{i=0}^{m-1}\beta_{i}^{\text{true}}\,B_{i}\left(z;t,k=2\right)+g^{\text{surf}}\left(z\right).\label{eq:true_model_spline}
\end{eqnarray}
We choose the surface profile $g^{\text{surf}}\left(z\right)$ and
use Equation (\ref{eq:psi_separable}) to obtain the starting model
of our supergranule, ie
\begin{equation}
\psi^{\text{start}}(\mathbf{x})=f\left(x\right)g^{\text{surf}}\left(z\right).
\end{equation}
The starting flow profile in our inversion is the same as the true flow above the surface, and falls to zero continuously just below. We can represent the vertical profile of the starting supergranule stream function in a basis of splines by setting the coefficients of B-splines below the surface to zero, ie. 
\begin{eqnarray}
g^{\text{surf}}\left(z\right) & = & \sum_{i=0}^{N-1}\beta_{i}^{\text{start}}B_{i}\left(z;t,k=2\right)\label{eq:spl_coeff_start},\\
\beta_{i}^{\text{start}} & = & \begin{cases}
0 & i<m\\
\beta_{i}^{\text{true}} & i\ge m
\end{cases}.
\end{eqnarray}

This is the model that we shall iteratively update, therefore at the first step of the inversion we set $\psi^\text{iter}=\psi^\text{start}$, or equivalently $\beta^\text{iter}_i=\beta^\text{start}_i\,\forall i$.
This choice is different from that made by \citet{Hanasoge2014} and \citet{Bhattacharya2016}, where the starting model had no flows. Note that we use the same set of knots $\{t\}$  to represent the inverted model as the ones that we had used to expand $g^\text{true}(z)$ in Equation (\ref{eq:basis_coeffs_all}). The inversion is carried out to obtain the spline coefficients $\{\beta^\text{iter}_i\}$ that lie below the solar surface.

We substitute the spline expansion of the iterated stream function in Equation (\ref{eq:kernel_psi})
to obtain kernels in spline space as
\begin{eqnarray}
\delta\chi & = & \sum_{i=0}^{m-1}\left[\int_{\odot}d\mathbf{x}\,K_{\psi}\left(\mathbf{x}\right)f\left(x\right)B_{i}\left(z;t,k=2\right)\right]\delta\beta^\text{iter}_{i}\nonumber \\
 & = & \sum_{i=0}^{m-1}K_{i}\,\delta\beta^\text{iter}_{i}.
\end{eqnarray}
The discrete kernels $K_{i}$ indicate the sensitivity of travel-times
to individual B-spline coefficients. We use the Broyden\textendash Fletcher\textendash Goldfarb\textendash Shanno algorithm
\citep[BFGS,][]{Nocedal2006}  to iteratively update our
model of the supergranule flow profile and reduce the travel-time
misfit.  

\begin{figure*}
\centering{}\includegraphics[scale=0.55]{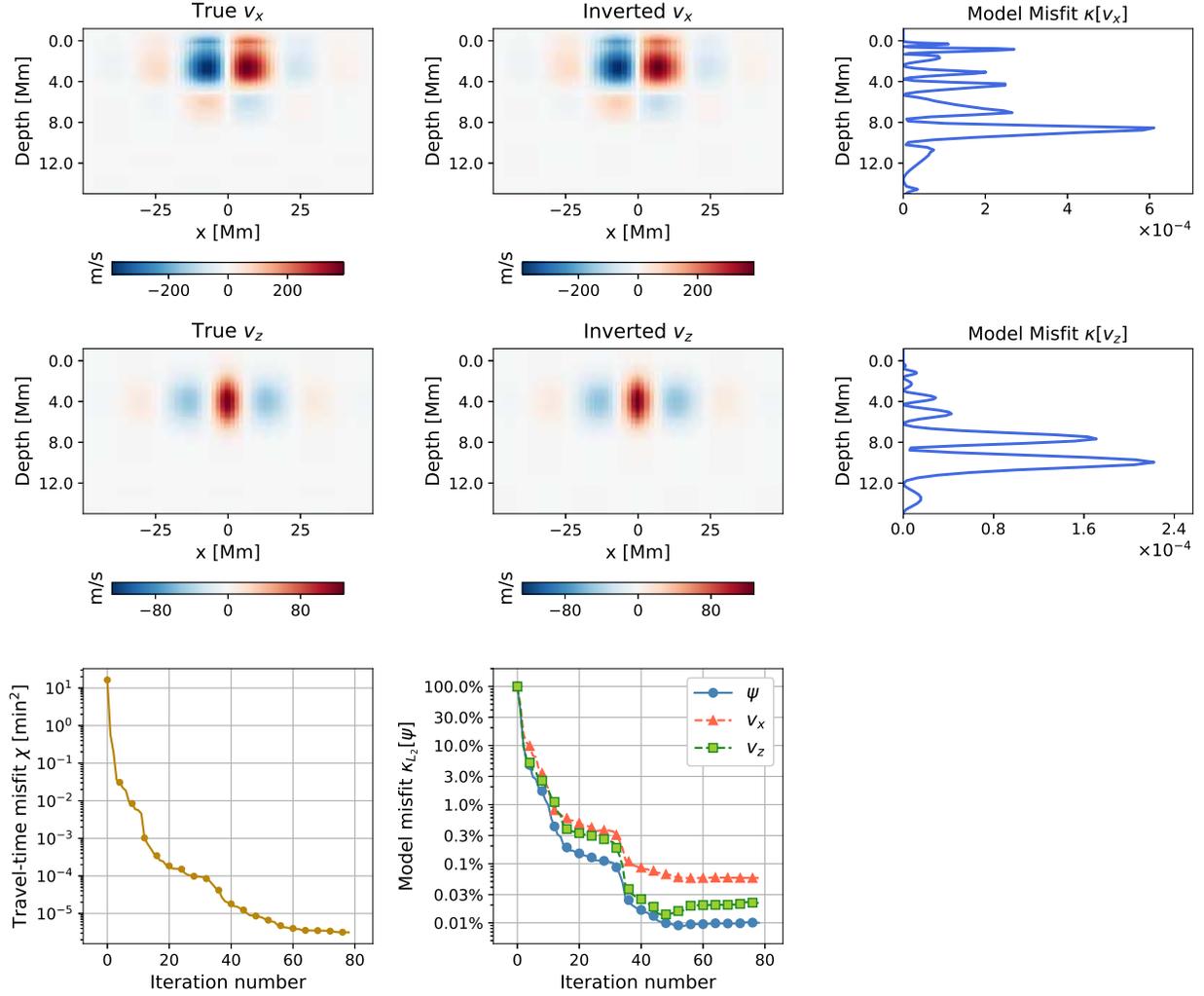}
\caption{\label{fig:misfits} True and inverted flow velocity for SG2
from Table \ref{tab:vertical_params}, with data and model misfits.
The panels are --- top left: true $v_{x}$, top center: inverted
$v_{x}$, top right: misfit in $v_{x}$ as a function of depth (Equation
\ref{eq:misfit_vx_z}), middle left: true $v_{z}$, middle center:
inverted $v_{z}$, middle right: misfit in $v_{z}$ as a function
of depth (Equation \ref{eq:misfit_vz_z}), bottom left: data misfit
from Equation (\ref{eq:travel-time-misfit}), bottom center: model
misfit from Equation (\ref{eq:model_misfit_L2}) for the stream function
$\psi$ and the two components of velocity. We see that the inverted
flow matches the true flow reasonably well, with the vertical
profiles differing by less than $0.5\%$.}
\end{figure*}
\begin{figure*}

\includegraphics[scale=0.5]{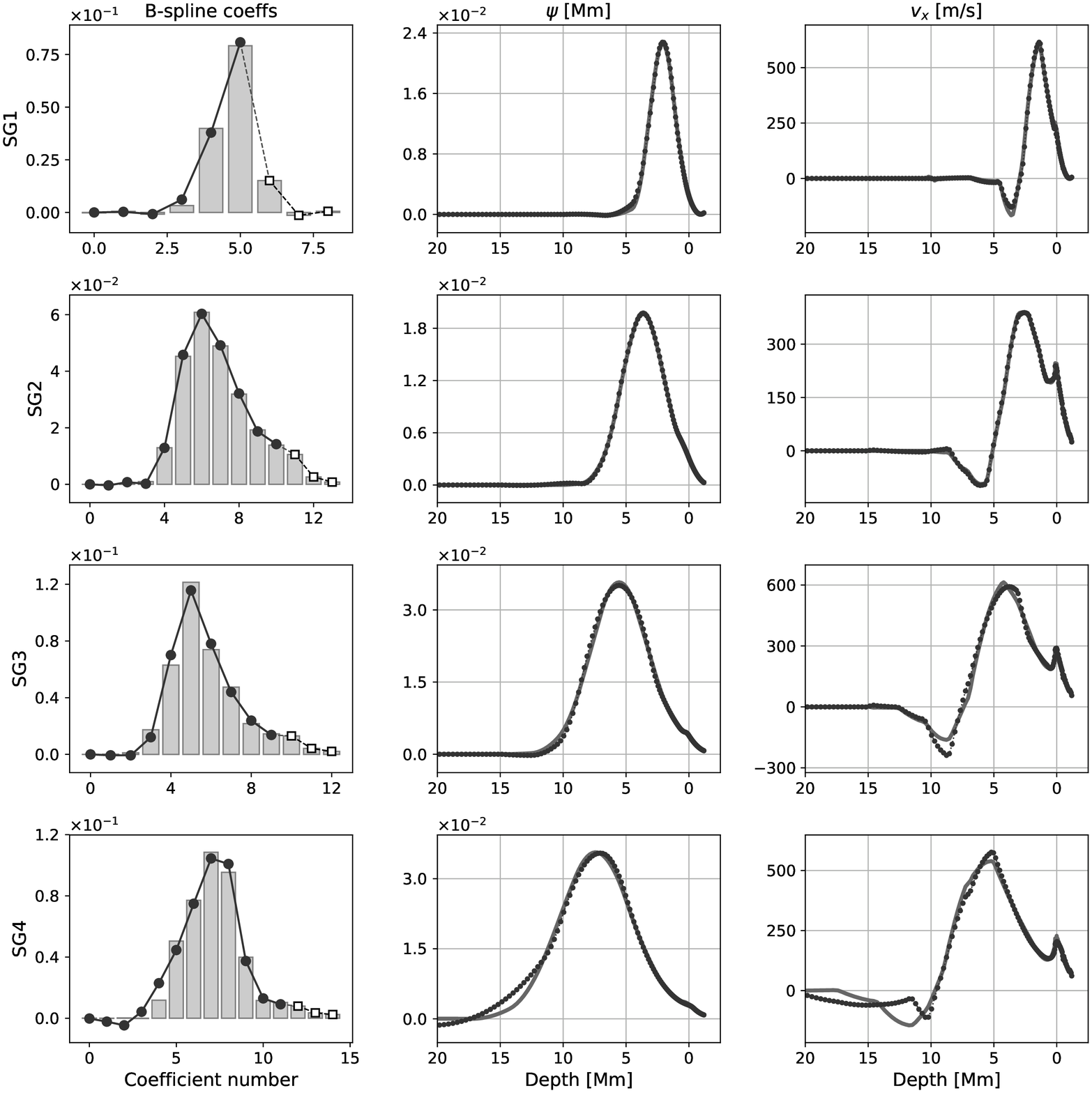}
\par

\caption{\label{fig:fits} True and inverted profiles for the four
supergranule Gaussian profiles from Table \ref{tab:vertical_params}.
Each row corresponds to one model, with the left panel representing
the B-spline coefficients, the middle panel showing the vertical profile of the stream function, and the rightmost panel depicting the vertical profile of the horizontal component of the flow velocity. The profile of vertical flow is not plotted, but is similar to the stream function. In the leftmost panel for each row, bars represent B-spline coefficients for the true stream function, white squares represent spline coefficients above the surface --- these are clamped to the value in the true model --- while black circles represent coefficients for the inverted solution. In the middle and right panels of each row, gray solid lines represents vertical profiles of the true models,
and black circles denote the profile for the inversion result. We find that
stream functions are reasonably well matched down to a depth of $10\,\text{Mm}$ 
from the surface, as expected for an inversion using $f-p_3$ modes.
The magnitude of horizontal return flow, however, is captured
correctly only the relatively shallower models SG1 and SG2.}
\end{figure*}
\begin{figure*}
\includegraphics[scale=0.6]{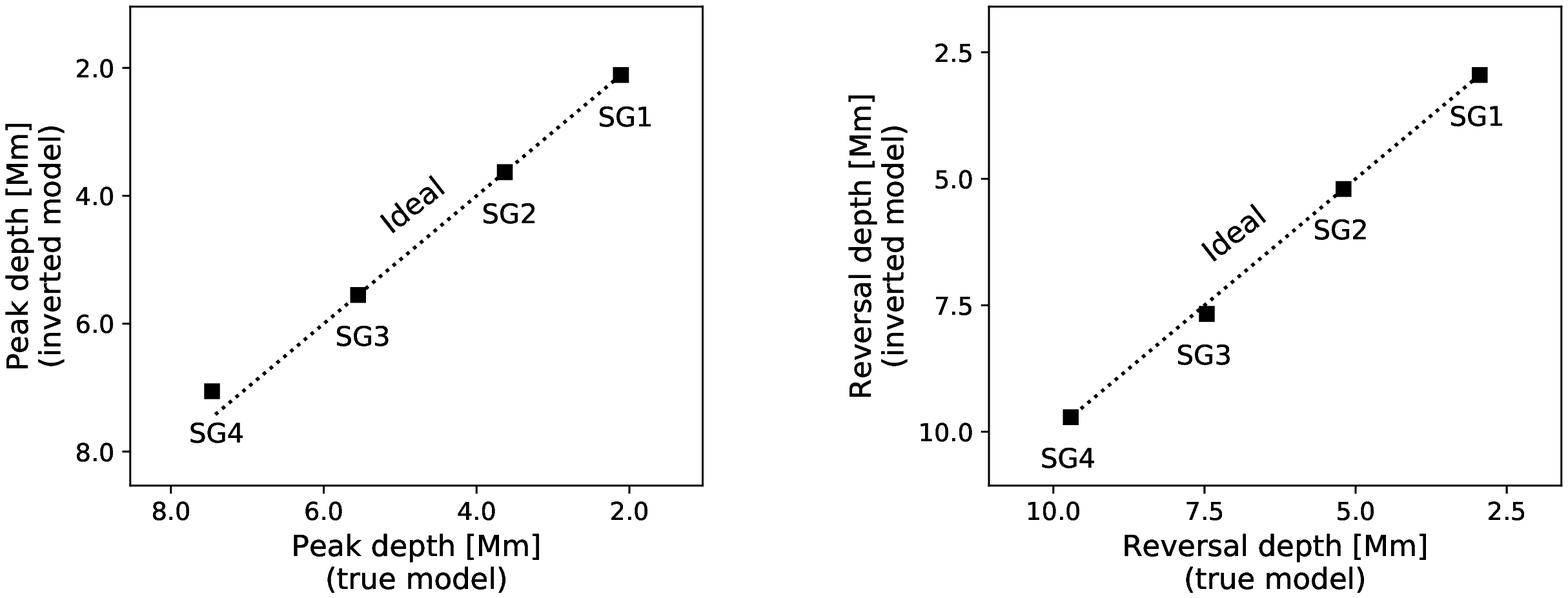}
\par

\caption{\label{fig:peak_depths}Left: Peak depths of the inverted stream function versus that of the true model, for each of the different supergranules from Table \ref{tab:vertical_params}. The dotted line indicates the ideal recovered peak depth, ie. that of the true model. 
Right: Reversal depths of $v_{x}$ in the inverted models against those in the true ones. The depth of reversal is one indicator of the vertical extent of a supergranule \citep{Duvall1998}.}
\end{figure*}

\subsection{Regularization}
The simulated spectrum (left panel in Figure \ref{fig:tt_modes}) features discernible modal ridges from $f$ to $p_{6}$, out of which we use ridges up to $p_{4}$ for our study. Restricting ourselves to a small set of modes imposes a limit on the depth until which we can infer flows, beyond this waves have limited sensitivity to flows. We therefore compute the knots required for the basis expansion in Equation (\ref{eq:basis_coeffs_all}) with a lower cutoff imposed. The depth of the cutoff is governed by the modes used in the inversion for each model, but it is chosen to be deep enough to ensure that the entire flow profile is contained within the spatial range.

We can estimate the depths that seismic waves probe by computing the asymptotic lower turning point \citep{Giles2000}. The turning-point for a wave that has the maximum power in the $p_{3}$ ridge --- corresponding to $k_xR_\odot=533$ and  temporal frequency $\nu=4.5\,\text{mHz}$ in the simulation --- is $9\,\text{Mm}$. We therefore expect inferences using radial orders $f-p_3$ to be accurate down to this depth. This also indicates that waves from the $p_4$ and higher radial orders might be necessary to infer flows deeper down. The value of lower cutoff and number of spline coefficients used for each model is listed in Table \ref{tab:number_of_coeffs} in Appendix \ref{app:spline}. We ensure that the iterated flow model falls smoothly to zero at the lower cutoff by multiplying the $i^{th}$ B-spline coefficient by a factor of $1/(1+\exp(-(i-i_2)/0.2))$, where $i_2$ is the index of the coefficient that lies $2$ Mm above the lower cutoff. For models SG1, SG2 and SG3 we obtain $i_2=1$, indicating that the two deepest coefficients $(i=0\;\text{and}\,i=1)$ are suppressed by factors of $150$ and $2$ respectively, whereas for SG4 we obtain $i_2=0$, indicating that the value of the deepest coefficient $(i=0)$ is reduced by a factor of $2$.

Previous analysis by \citet{Bhattacharya2016} had used spatial smoothing to reduce high spatial-frequency variation in the numerically computed sensitivity kernel. This is not strictly necessary in our approach, and we found minor differences by including smoothing.

\section{Results and discussion}

The ``inversion'' in our analysis is a series of forward simulations,
followed by optimization in the parameter space of the flow. Each
stage in the iterative optimization proceeds by reducing the travel-time
misfit in Equation (\ref{eq:travel-time-misfit}). We plot the travel-time
for different radial orders for the model SG2 in Figure \ref{fig:tt_modes}.
We compare the travel-time shift with the horizontal flow that is
indicated by the colored patch. The travel-time shift is a measure
how much the waveform in the starting model is delayed with respect
to that in the true model, negative values indicating that the
wavepacket in the starting model arrives earlier at a receiver in
relation to the true model.

Each model is updated by iteratively reducing the travel-time misfit
using Equation (\ref{eq:kernel_psi}). In solar travel-time measurements,
error bars arising form realization noise provide a natural stopping
point for iterations; in the absence of noise we iterate until the
relative change in travel-misfit falls below $0.1\%$. We quantify
the efficacy of the inversion by defining model misfits for the stream
function and the components of the flow. The flow velocities are related
to derivatives of stream function through Equation (\ref{eq:v_psi}),
so the misfit in components of flow, when computed at each depth depth,
differs from that for the stream function. The misfit $\kappa$ for
each parameter is defined as the normalized square of the difference
between true and iterated models evaluated as a function of depth
by averaging over the horizontal direction $x$, ie. for the stream function $\psi$ we obtain
\begin{equation}
\kappa\left[\psi\right]\left(z\right)=\frac{\int dx\left(\psi^{\text{true}}\left(x,z\right)-\psi^{\text{iter}}\left(x,z\right)\right)^{2}}{\int dx\,\psi^{\text{true}}\left(x,z=\text{argmax}\,g^{\text{true}}\left(z\right)\right)^{2}},
\end{equation}
where the horizontal integral in the denominator is evaluated at the
depth where the true model reaches its peak. This ensures that the maximum
misfit for the starting model is normalized to one. The notation here
indicates that the misfit $\kappa$ is computed for the parameter in square brackets,
and is evaluated as a function of vertical layer $z$. Using the separability
condition in Equation (\ref{eq:psi_separable}), we can express $\kappa\left[\psi\right]$
in term of the vertical profile $g\left(z\right)$ as 
\begin{equation}
\kappa\left[\psi\right]\left(z\right)=\frac{\left(g^{\text{true}}\left(z\right)-g^{\text{iter}}\left(z\right)\right)^{2}}{\left[\max g^{\text{true}}\left(z\right)\right]^{2}}.\label{eq:misfit_psi_z}
\end{equation}
We define analogous misfit functions for the two components of flow
velocity, the expressions being 

\begin{eqnarray}
\kappa\left[v_{x}\right]\left(z\right) & = & \frac{\left[\frac{1}{\rho}\partial_{z}\left(\rho c\left(g^{\text{true}}\left(z\right)-g^{\text{iter}}\left(z\right)\right)\right)\right]^{2}}{\left[\max\left(\frac{1}{\rho}\partial_{z}\left(\rho c\,g^{\text{true}}\left(z\right)\right)\right)\right]^{2}}.\label{eq:misfit_vx_z}\\
\kappa\left[v_{z}\right]\left(z\right) & = & \frac{c^{2}\left(g^{\text{true}}\left(z\right)-g^{\text{iter}}\left(z\right)\right)^{2}}{\left[\max\left(c\,g^{\text{true}}\left(z\right)\right)\right]^{2}}.\label{eq:misfit_vz_z}
\end{eqnarray}

Alongside studying misfit as a function of depth, we consider the
normalized $L_{2}$ norm of differences between true and iterated
models integrated over the entire space, defined as 
\begin{equation}
\kappa_{L_{2}}\left[\psi\right]\left(z\right)=\frac{\int dz\left(g^{\text{true}}\left(z\right)-g^{\text{iter}}\left(z\right)\right)^{2}}{\int dz\left(g^{\text{true}}\left(z\right)-g^{\text{start}}\left(z\right)\right)^{2}},\label{eq:model_misfit_L2}
\end{equation}
to gain insight into the degree of improvement to the model after
each iteration. We compare the inverted flow velocity field with the profile of the model SG2 in Figure \ref{fig:misfits},
and analyze the model misfit. We find that the inversion result matches
the true model reasonably well, with the stream function misfit
$\kappa_{L_{2}}\left[\psi\right]$ being around $0.01\%$ after the final iteration.

We plot the inverted vertical profiles of all the models in Figure
\ref{fig:fits}. One question we ask in this paper is whether seismic
waves can estimate the depth of supergranules. We plot the expected
and inferred peak depths for each model in Figure \ref{fig:peak_depths}.
We find that we recover the peak depths accurately for all the models.
It is encouraging to note that we are able to extract Gaussian profiles
up to a depth of $8-10\,\text{Mm}$, that is beyond the $6\,\text{Mm}$ limit
found by previous seismic inferences in the presence
of realization noise \citep{Braun2007,Woodard2007}. We were however
unable to retrieve the profile for a deeper model with a peak depth
of $14\,\text{Mm}$ with $f-p_{4}$ modes. It might be interesting
to see if the introduction of higher $p$ modes makes a difference. 

Aside from peak depth, another important parameter that is used to estimate 
supergranules depth is the layer at
which the horizontal flow reverses direction. All the models that
we study have a reversal in $v_{x}$, keeping with the assumption
of a steady convective cell. We compare the true and inferred reversal
depths in Figure \ref{fig:peak_depths}. We find that the inversion
reproduces comparable values, although the exact profile of the return
flow is not accurately captured. Whether the reversal actually takes place 
is subject to debate, since it is not detected in the study by \citet{Woodard2007} 
and suggested to be spurious by \citet{Svanda2013ApJ...775....7S,Degrave2014}. Our results seem to indicate that the depth would be captured correctly for shallow models if the flow does reverse, provided the inference is not limited by noise.

In this work we have used iterative forward modeling to obtain the
best-fit flow model given travel-time measurements at the solar surface.
This approach is inherently non-linear, as it requires re-computing
the sensitivity kernel after each iteration. From Figure \ref{fig:misfits},
we see that a substantial drop in model misfit takes place in the
first few iterations. This makes it interesting to compare this approach
with a linear inversion; given the small parameter space, it might
be possible to pinpoint the differences in the inverted flow arising
from the two approaches. We have also not considered any noise associated
with travel-time measurements, so this analysis can not be directly
applied to solar measurements. It would be interesting to study the
extent to which the inferences are affected in presence of noise,
and develop a slightly modified technique that accounts for realistic
nose covariance matrices \citep{Gizon2004,Svanda2011}. It was also
pointed out by \citet{Degrave2014} that validation tests with frozen
non-magnetic flow fields might be too idealized a scenario compared
to studying a Doppler time-series obtained from the Sun. Therefore
an extension of this work to flows present in realistic solar simulations
might be in order.

\begin{acknowledgements}

SMH acknowledges support from Ramanujan fellowship SB/S2/RJN-73/2013,
the Max-Planck partner group program and thanks the Center for Space
Science, New York University at Abu Dhabi. JB acknowledges the financial
support provided by the Department of Atomic Energy, India.

\end{acknowledgements}

\bibliographystyle{aa}
\bibliography{references}

\begin{thebibliography}{44}
\expandafter\ifx\csname natexlab\endcsname\relax\def\natexlab#1{#1}\fi

\bibitem[{{Bhattacharya} \& {Hanasoge}(2016)}]{Bhattacharya2016}
{Bhattacharya}, J. \& {Hanasoge}, S.~M. 2016, \apj, 826, 105

\bibitem[{{Braun} {et~al.}(2007){Braun}, {Birch}, {Benson}, {Stein}, \&
  {Nordlund}}]{Braun2007}
{Braun}, D.~C., {Birch}, A.~C., {Benson}, D., {Stein}, R.~F., \& {Nordlund},
  {\AA}. 2007, \apj, 669, 1395

\bibitem[{{Braun} {et~al.}(2004){Braun}, {Birch}, \& {Lindsey}}]{Braun2004}
{Braun}, D.~C., {Birch}, A.~C., \& {Lindsey}, C. 2004, in ESA Special
  Publication, Vol. 559, SOHO 14 Helio- and Asteroseismology: Towards a Golden
  Future, ed. D.~{Danesy}, 337

\bibitem[{{Christensen-Dalsgaard} {et~al.}(1996){Christensen-Dalsgaard},
  {Dappen}, {Ajukov}, {Anderson}, {Antia}, {Basu}, {Baturin}, {Berthomieu},
  {Chaboyer}, {Chitre}, {Cox}, {Demarque}, {Donatowicz}, {Dziembowski},
  {Gabriel}, {Gough}, {Guenther}, {Guzik}, {Harvey}, {Hill}, {Houdek},
  {Iglesias}, {Kosovichev}, {Leibacher}, {Morel}, {Proffitt}, {Provost},
  {Reiter}, {Rhodes}, {Rogers}, {Roxburgh}, {Thompson}, \&
  {Ulrich}}]{ChristensenDalsgaard1996}
{Christensen-Dalsgaard}, J., {Dappen}, W., {Ajukov}, S.~V., {et~al.} 1996,
  Science, 272, 1286

\bibitem[{{DeGrave} {et~al.}(2014){DeGrave}, {Jackiewicz}, \&
  {Rempel}}]{Degrave2014}
{DeGrave}, K., {Jackiewicz}, J., \& {Rempel}, M. 2014, \apj, 788, 127

\bibitem[{Dierckx(1993)}]{Dierckx1993}
Dierckx, P. 1993, Curve and Surface Fitting with Splines (New York, NY, USA:
  Oxford University Press, Inc.)

\bibitem[{Dombroski {et~al.}(2013)Dombroski, Birch, Braun, \&
  Hanasoge}]{Dombroski2013}
Dombroski, D.~E., Birch, A.~C., Braun, D.~C., \& Hanasoge, S.~M. 2013, Solar
  Physics, 282, 361

\bibitem[{Duvall \& Birch(2010)}]{Duvall2010}
Duvall, J. \& Birch, A.~C. 2010, \apj, 725, L47

\bibitem[{Duvall \& Hanasoge(2012)}]{Duvall2012}
Duvall, T. \& Hanasoge, S. 2012, \solphys, 136

\bibitem[{{Duvall} {et~al.}(2014){Duvall}, {Hanasoge}, \&
  {Chakraborty}}]{Duvall2014}
{Duvall}, T.~L., {Hanasoge}, S.~M., \& {Chakraborty}, S. 2014, \solphys, 289,
  3421

\bibitem[{{Duvall}(1998)}]{Duvall1998}
{Duvall}, Jr., T.~L. 1998, in ESA Special Publication, Vol. 418, Structure and
  Dynamics of the Interior of the Sun and Sun-like Stars, ed. S.~{Korzennik},
  581

\bibitem[{{Duvall} \& {Gizon}(2000)}]{Duvall2000}
{Duvall}, Jr., T.~L. \& {Gizon}, L. 2000, \solphys, 192, 177

\bibitem[{{Duvall Jr.} {et~al.}(1993){Duvall Jr.}, Jefferies, Harvey, \&
  Pomerantz}]{Duvall1993}
{Duvall Jr.}, T., Jefferies, S., Harvey, J., \& Pomerantz, M. 1993, \nat, 362,
  430

\bibitem[{{Giles}(2000)}]{Giles2000}
{Giles}, P.~M. 2000, PhD thesis, STANFORD UNIVERSITY

\bibitem[{{Gizon} \& {Birch}(2002)}]{Gizon2002}
{Gizon}, L. \& {Birch}, A.~C. 2002, \apj, 571, 966

\bibitem[{{Gizon} \& {Birch}(2004)}]{Gizon2004}
{Gizon}, L. \& {Birch}, A.~C. 2004, \apj, 614, 472

\bibitem[{{Gizon} {et~al.}(2010){Gizon}, {Birch}, \& {Spruit}}]{Gizon2010}
{Gizon}, L., {Birch}, A.~C., \& {Spruit}, H.~C. 2010, \araa, 48, 289

\bibitem[{{Gizon} {et~al.}(2003){Gizon}, {Duvall}, \& {Schou}}]{Gizon2003}
{Gizon}, L., {Duvall}, T.~L., \& {Schou}, J. 2003, \nat, 421, 43

\bibitem[{{Gizon} {et~al.}(2000){Gizon}, {Duvall}, \& {Larsen}}]{Gizon2000}
{Gizon}, L., {Duvall}, Jr., T.~L., \& {Larsen}, R.~M. 2000, Journal of
  Astrophysics and Astronomy, 21, 339

\bibitem[{Hanasoge(2014)}]{Hanasoge2014}
Hanasoge, S.~M. 2014, \apj, 797, 23

\bibitem[{{Hanasoge} {et~al.}(2011){Hanasoge}, {Birch}, {Gizon}, \&
  {Tromp}}]{Hanasoge2011}
{Hanasoge}, S.~M., {Birch}, A., {Gizon}, L., \& {Tromp}, J. 2011, \apj, 738,
  100

\bibitem[{{Hanasoge} \& {Duvall}(2007)}]{Hanasoge2007}
{Hanasoge}, S.~M. \& {Duvall}, Jr., T.~L. 2007, Astronomische Nachrichten, 328,
  319

\bibitem[{{Hanasoge} {et~al.}(2006){Hanasoge}, {Larsen}, {Duvall}, {De Rosa},
  {Hurlburt}, {Schou}, {Roth}, {Christensen-Dalsgaard}, \&
  {Lele}}]{Hanasoge2006}
{Hanasoge}, S.~M., {Larsen}, R.~M., {Duvall}, Jr., T.~L., {et~al.} 2006, \apj,
  648, 1268

\bibitem[{{Hathaway} {et~al.}(2000){Hathaway}, {Beck}, {Bogart}, {Bachmann},
  {Khatri}, {Petitto}, {Han}, \& {Raymond}}]{Hathaway2000}
{Hathaway}, D.~H., {Beck}, J.~G., {Bogart}, R.~S., {et~al.} 2000, \solphys,
  193, 299

\bibitem[{{Hu} {et~al.}(1996){Hu}, {Hussaini}, \& {Manthey}}]{Hu96}
{Hu}, F.~Q., {Hussaini}, M.~Y., \& {Manthey}, J.~L. 1996, Journal of
  Computational Physics, 124, 177

\bibitem[{{Jackiewicz} {et~al.}(2008){Jackiewicz}, {Gizon}, \&
  {Birch}}]{Jackiewicz2008}
{Jackiewicz}, J., {Gizon}, L., \& {Birch}, A.~C. 2008, \solphys, 251, 381

\bibitem[{{Jackiewicz} {et~al.}(2007){Jackiewicz}, {Gizon}, {Birch}, \&
  {Duvall}}]{Jackiewicz2007}
{Jackiewicz}, J., {Gizon}, L., {Birch}, A.~C., \& {Duvall}, Jr., T.~L. 2007,
  \apj, 671, 1051

\bibitem[{{Langfellner} {et~al.}(2015){Langfellner}, {Gizon}, \&
  {Birch}}]{Langfellner2015}
{Langfellner}, J., {Gizon}, L., \& {Birch}, A.~C. 2015, \aap, 579, L7

\bibitem[{{Lele}(1992)}]{Lele92}
{Lele}, S.~K. 1992, Journal of Computational Physics, 103, 16

\bibitem[{Nocedal \& Wright(2006)}]{Nocedal2006}
Nocedal, J. \& Wright, S.~J. 2006, Numerical Optimization, 2nd edn. (New York:
  Springer)

\bibitem[{{Nordlund} {et~al.}(2009){Nordlund}, {Stein}, \&
  {Asplund}}]{Nordlund2009}
{Nordlund}, {\AA}., {Stein}, R.~F., \& {Asplund}, M. 2009, Living Reviews in
  Solar Physics, 6, 2

\bibitem[{{Rieutord} {et~al.}(2008){Rieutord}, {Meunier}, {Roudier}, {Rondi},
  {Beigbeder}, \& {Par{\`e}s}}]{Rieutord2008}
{Rieutord}, M., {Meunier}, N., {Roudier}, T., {et~al.} 2008, \aap, 479, L17

\bibitem[{{Rieutord} {et~al.}(2010){Rieutord}, {Roudier}, {Rincon}, {Malherbe},
  {Meunier}, {Berger}, \& {Frank}}]{Rieutord2010}
{Rieutord}, M., {Roudier}, T., {Rincon}, F., {et~al.} 2010, \aap, 512, A4

\bibitem[{{Schou} {et~al.}(2012){Schou}, {Scherrer}, {Bush}, {Wachter},
  {Couvidat}, {Rabello-Soares}, {Bogart}, {Hoeksema}, {Liu}, {Duvall}, {Akin},
  {Allard}, {Miles}, {Rairden}, {Shine}, {Tarbell}, {Title}, {Wolfson},
  {Elmore}, {Norton}, \& {Tomczyk}}]{Schou2012}
{Schou}, J., {Scherrer}, P.~H., {Bush}, R.~I., {et~al.} 2012, \solphys, 275,
  229

\bibitem[{{Stein} \& {Nordlund}(2001)}]{Stein2001}
{Stein}, R.~F. \& {Nordlund}, {\AA}. 2001, \apj, 546, 585

\bibitem[{{Tromp} {et~al.}(2010){Tromp}, {Luo}, {Hanasoge}, \&
  {Peter}}]{Tromp2010}
{Tromp}, J., {Luo}, Y., {Hanasoge}, S., \& {Peter}, D. 2010, Geophysical
  Journal International, 183, 791

\bibitem[{{{\v S}vanda}(2012)}]{Svanda2012}
{{\v S}vanda}, M. 2012, \apjl, 759, L29

\bibitem[{{{\v S}vanda}(2013)}]{Svanda2013ApJ...775....7S}
{{\v S}vanda}, M. 2013, \apj, 775, 7

\bibitem[{{{\v S}vanda}(2015)}]{Svanda2015}
{{\v S}vanda}, M. 2015, \aap, 575, A122

\bibitem[{{{\v S}vanda} {et~al.}(2011){{\v S}vanda}, {Gizon}, {Hanasoge}, \&
  {Ustyugov}}]{Svanda2011}
{{\v S}vanda}, M., {Gizon}, L., {Hanasoge}, S.~M., \& {Ustyugov}, S.~D. 2011,
  \aap, 530, A148

\bibitem[{{{\v S}vanda} {et~al.}(2013){{\v S}vanda}, {Roudier}, {Rieutord},
  {Burston}, \& {Gizon}}]{Svanda2013}
{{\v S}vanda}, M., {Roudier}, T., {Rieutord}, M., {Burston}, R., \& {Gizon}, L.
  2013, \apj, 771, 32

\bibitem[{{Woodard}(2007)}]{Woodard2007}
{Woodard}, M.~F. 2007, \apj, 668, 1189

\bibitem[{{Zhao}(2004)}]{Zhao2004}
{Zhao}, J. 2004, PhD thesis, STANFORD UNIVERSITY

\bibitem[{{Zhao} \& {Kosovichev}(2003)}]{Zhao2003}
{Zhao}, J. \& {Kosovichev}, A.~G. 2003, in ESA Special Publication, Vol. 517,
  GONG+ 2002. Local and Global Helioseismology: the Present and Future, ed.
  H.~{Sawaya-Lacoste}, 417--420

\end{thebibliography}

\begin{appendix}

\section{Travel-time measurements\label{ttsec}}

\begin{figure*}

\includegraphics[scale=0.55]{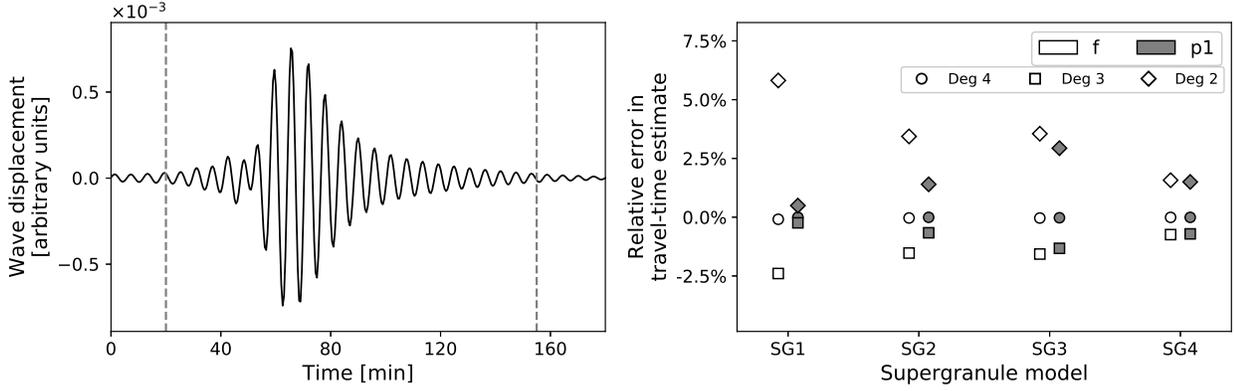}

\caption{\label{fig:travel_time_validation}
Left: One example of an $f-$mode wavepacket measured at a receiver located at $x_{\text{r}}=  30\,\text{Mm}$ for the supergranule SG1. The extent of the window function 
$w\left(x_{\text{r}},t\right)$ is denoted by vertical dashed lines.
Right: Relative error in estimated travel-time
shifts using wavepackets corresponding to radial orders $f$ and $p_{1}$ (indicated by different colors) as a function of different degree of truncation of Equation (\ref{eq:disp_misfit_expanded})
(indicated by different symbols), for different supergranule models.
Truncating Equation (\ref{eq:disp_misfit_expanded}) to quadratic order results in a linear relation between travel-time shift $\delta\tau$ and wave displacement $\xi^{\text{iter}}$, this is plotted with diamonds.}

\end{figure*}

We measure travel-time shifts by minimizing the squared difference between wave displacements in the true and starting supergranule simulations. Given wave displacements $\xi^{\text{true}}$ and $\xi^{\text{iter}}$
recorded at a receiver at $x_{\text{r}}$ and filtered to obtain the wavepacket corresponding to a specific radial order, we define a misfit 
\begin{eqnarray}
\eta\left(x_{\text{r}},\tau\right) & = & \int dt\,w\left(x_{\text{r}},t\right)\times\nonumber \\
 &  & \left(\xi^{\text{iter}}\left(x_{\text{r}},t\right)-\xi^{\text{true}}\left(x_{\text{r}},t-\tau\right)\right)^{2}\label{eq:receiver_disp_misfit}
\end{eqnarray}
where $w\left(x_{\text{r}},t\right)$ is a window function that encloses the wavepackets and isolates them from artifacts that might arise from spatio-temporal periodicity assumed in the simulation.
Specifically, we choose the functional form of $w\left(x_{\text{r}},t\right)$
to be a box function, that is one over the range of the wavepacket
and zero outside. We plot one example of a measured wavepacket and window function in Figure \ref{fig:travel_time_validation} (left panel).
The travel-time shift is defined as the value of
$\tau$ that minimizes $\eta\left(x_{\text{r}},\tau\right)$ \citep{Gizon2002,Gizon2004}.
For band-limited waveforms sampled beyond twice their Nyquist frequency,
this can be computed by equating the temporal derivative of $\eta\left(x_{\text{r}},\tau\right)$
to zero and solving for $\tau$. Expanding $\eta$ in terms of the
time-shift $\tau$, suppressing the explicit dependence on the coordinates
$x_{\text{r}}$ and $t$, representing the order of derivative using
superscripts in parentheses and referring to $\xi^{\text{true}}\left(x_{\text{r}},t\right)-\xi^{\text{iter}}\left(x_{\text{r}},t\right)$
as $\delta\xi$, we obtain

\begin{flalign}
\eta\left(x_{\text{r}},\tau\right) & =\int dt\,w\left(x_{\text{r}},t\right)\left(\delta\xi\right)^{2}\nonumber \\
 & -\tau\int dt\,w\left(x_{\text{r}},t\right)2\delta\xi\,\xi^{\text{true}}{}^{\left(1\right)}\nonumber \\
 & +\tau^{2}\int dt\,w\left(x_{\text{r}},t\right)\left(\delta\xi\,\xi^{\text{true}}{}^{\left(2\right)}+\left(\xi^{\text{true}}{}^{\left(1\right)}\right)^{2}\right)\nonumber \\
 & +\mathcal{O}\left(\tau^{3}\right).\label{eq:disp_misfit_expanded}
\end{flalign}

Retaining terms till quadratic order and solving $\partial_{\tau}\eta\left(x_{\text{r}},\tau\right)=0$
leads to a travel-time shift given by
\begin{eqnarray}
\delta\tau\left(x_{\text{r}}\right) & = & \frac{\int dt\,w\left(x_{\text{r}},t\right)\,\xi^{\text{true}}{}^{\left(1\right)}\,\delta\xi}{\int dt\,w\left(x_{\text{r}},t\right)\left(\delta\xi\,\xi^{\text{true}}{}^{\left(2\right)}+\left(\xi^{\text{true}}{}^{\left(1\right)}\right)^{2}\right)}\nonumber \\
 & \approx & \int dt\,\left[\frac{w\left(x_{\text{r}},t\right)\xi^{\text{true}}{}^{\left(1\right)}}{\int dt^{\prime}\,w\left(x_{\text{r}},t^{\prime}\right)\left(\xi^{\text{true}}{}^{\left(1\right)}\right)^{2}}\right]\,\delta\xi.\label{eq:tt2}
\end{eqnarray}
Equation (\ref{eq:tt2}) is in the form of travel-time shift defined
by \citet{Gizon2002}, where the shift $\delta\tau$ is linear in
the displacement difference $\delta\xi$. The linear dependence of
travel-time shifts on $\delta\xi$ is important for consistent computation
of sensitivity kernels in the first Born approximation. At high flow
velocities, however, it is possible that this linear relationship
fails to remain a good approximation \citep{Jackiewicz2007,Degrave2014}.
For seismic waves with frequency $\omega$, the deviation from linearity
will be noticeable if the measured travel-time shift $\delta\tau$
satisfies $\left|\omega\,\delta\tau\right|\ll1$. We have carried out validation tests for by artificially time-shifting a simulated wave field by
a typical value measured at surface for each supergranule model, and
trying to recover the shift from the series expansion of $\eta\left(x_{\text{r}},\tau\right)$ truncated at various orders in $\tau$. We plot the result of one such validation test in Figure \ref{fig:travel_time_validation} (right panel).
We find that travel-times computed using the linear approximation
for the various supergranule models are within $5\%$ of the expected
value, the accuracy of the estimate improving with an increase in
degree of the truncation. Note that this is the error in travel-time measurement at the first iteration in our inversion. Subsequent iterations improve the flow model and lead to a significant decrease in $\delta\tau$, consequently the error in measuring travel-time shifts is also reduced. This ensures that the inverted flow is not affected by the error in measured travel time for the starting model. It might be necessary to take this error into consideration for a linear inversion. 

\section{Spline expansion\label{app:spline}}

\begin{figure*}
\includegraphics[scale=0.61]{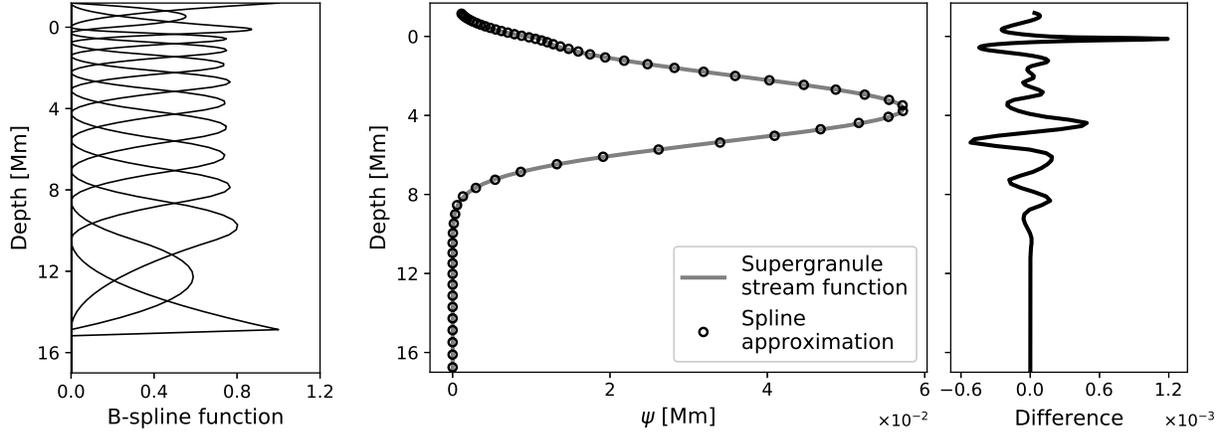}

\caption{\label{fig:splines}Left: B-spline functions used in the expansion of the supergranule model SG2. Specific choice of knots might be important for the inversion to converge to the correct model. Density of knots with depth is reflective of the degree of stratification. Note that the B-splines are not normalized by the corresponding coefficients. Middle: Supergranule model from Equation (\ref{eq:superganule_psi}) (grey solid line) and the smoothing spline approximation to it (black circles). Right: Error in the spline approximation with depth. We see that the smoothing spline approximation is fairly representative of the form of the vertical profile of the supergranule stream function.
}

\end{figure*}

We expand the supergranule model from Equation (\ref{eq:superganule_psi}) in a basis of B-splines following Equation (\ref{eq:basis_coeffs_all}) to obtain a set of coefficients that we invert for. The first step to this expansion is to obtain a set of knots that govern the B-splines. The choice of knots might be important for the inversion
to succeed, however further study needs to be carried out to establish this. We choose knots by applying the Dierckx algorithm \citep{Dierckx1993}
to the true supergranule, but alternate choices of knots derived
from an independent, vertically stratified parameter such as sound-speed may also be used. The knots and coefficients are computed using the
``scipy.interpolate'' package of the python programming language, that internally calls the Fortran library FITPACK. The module computes the spline fit by evaluating an optimal set of knots and coefficients, ensuring that the squared $L_2$ norm of the difference between the data being fit and the spline approximant falls below a specified smoothing factor. Reducing the value of this factor improves the fit; this is achieved by updating the set of knots followed by reevaluating the expansion coefficients. We carry out our inversion in a space spanned by the spline coefficients, therefore the specific choice of a approximant is a tradeoff between the quality of fit and the number of parameters used to obtain the fit. This is why we choose smoothing splines over interpolating ones, since the latter involves a similar fit evaluated without any smoothing and produces a set of knots similar in size to the number of grid points, while the former can be tuned to significantly trim down the size of this set. We select the smoothing factor through experimentation to obtain an accurate representation of the stream function in the B-spline basis while restraining the number of coefficients to around $10$. We list the smoothing parameters and knots for each of the supegranule models in Table \ref{tab:number_of_coeffs}. We decide upon quadratic splines as a compromise between a cubic-spline fit that is more oscillatory and a linear-spline fit that is not as smooth and results in a larger parameter space.

We plot the functional form of the B-spline functions for model SG2 in Figure \ref{fig:splines} (left panel). We plot the functional form of the supergranule from Equation (\ref{eq:basis_coeffs_all}) as well as the spline approximation to it in Figure \ref{fig:splines} (middle panel), and we plot the error in the approximation in the right panel. We find that smoothing splines are a reasonably good representation of this supergranule model beneath the solar surface.

\begin{table*}
\caption{\label{tab:number_of_coeffs}B-spline parameters used in the inversion}
\renewcommand{\arraystretch}{1.2}
\begin{tabular}{ c c c c c }

\hline 

\multirow{3}{*}{Model} & \multicolumn{1}{p{2cm} }{\centering Depth of \\lower cutoff} & \multicolumn{1}{p{3cm} }{\centering Smoothing \\parameter} & \multicolumn{1}{p{2cm} }{\centering Number of \\coefficients}  & \multirow{2}{*}{Knots}\tabularnewline
& [Mm] & [Mm$^2$] & & [Mm] \tabularnewline
\hline 
\hline 
\multirow{2}{*}{SG1} & \multirow{2}{*}{$10$}  & \multirow{2}{*}{$3.6\times 10^{-5}$} & \multirow{2}{*}{$9$} &
\multicolumn{1}{p{7cm}}{-10.0, -10.0, -10.0, -6.9, -4.5, -3.5, -2.7, -1.3, 0.2, 1.2, 1.2, 1.2}\tabularnewline
\hline 
\multirow{2}{*}{SG2} & \multirow{2}{*}{$15$}  & \multirow{2}{*}{$7.3 \times 10^{-6}$} & \multirow{2}{*}{$14$} &
\multicolumn{1}{p{7cm}}{-14.9, -14.9, -14.9, -10.5, -8.5, -7.1, -5.5, -4.2, -3.1, -2.2, -1.4, -0.8, -0.3, -0.0, 1.2, 1.2, 1.2}\tabularnewline
\hline 
\multirow{2}{*}{SG3} & \multirow{2}{*}{$15$} & \multirow{2}{*}{$2.8 \times 10^{-4}$} & \multirow{2}{*}{$13$}  &
\multicolumn{1}{p{7cm}}{-14.9, -14.9, -14.9, -12.6, -10.5, -8.5, -7.1, -4.2, -3.1, -2.2, -0.8, -0.3, -0.0, 1.2, 1.2, 1.2}\tabularnewline
\hline 
\multirow{2}{*}{SG4} & \multirow{2}{*}{$30$} & \multirow{2}{*}{$1.6 \times 10^{-5}$} & \multirow{2}{*}{$15$}   &
\multicolumn{1}{p{7cm}}{-29.7, -29.7, -29.7, -20.9, -17.4, -14.3, -11.5, -10.2, -9.0, -6.9, -5.0, -2.1, -0.5, -0.0, 0.4, 1.2, 1.2, 1.2}\tabularnewline
\hline 
\end{tabular}

\end{table*}

\end{appendix}

\end{document}